\documentclass[a4paper,fleqn,useAMS,usenatbib]{mnras}

\pdfoutput=1

\usepackage{mathptmx}

\usepackage[T1]{fontenc}
\usepackage{ae,aecompl}

\usepackage[dvips]{graphicx}
\usepackage{amsmath}
\usepackage{amsfonts}
\usepackage{amssymb}
\usepackage{comment}
\usepackage[dvipsnames]{xcolor}
\usepackage[%
        ]{hyperref}
        
\hypersetup{
	colorlinks=true,	
	urlcolor=MidnightBlue,		
	pdfpagelabels=true,
	hypertexnames=true,
	plainpages=false,
	naturalnames=true,
	pdftitle={Transit Clairvoyance},    
	pdfauthor={Kipping \& Lam},     
	linkcolor=WildStrawberry,          
	citecolor=ForestGreen,        
}

\newcommand{\KeplerMission}{ {\it Kepler Mission} }
\newcommand{\Kepler}{{\it Kepler}}
\newcommand{\TESS}{{\it TESS}}

\newcommand{\python}{{\tt PYTHON }}

\newcommand{\wwwcoolworlds}{\href{https://github.com/cl3425/Clairvoyance/}{this URL}}

\title[Transit Clairvoyance]{
Transit Clairvoyance:\\
Enhancing TESS follow-up using artificial neural networks
}
\author[Kipping \& Lam]{David M. Kipping$^{1}$\thanks{E-mail:
\href{mailto:dkipping@astro.columbia.edu}{dkipping@astro.columbia.edu}} and Christopher Lam$^{1}$\\
$^{1}$Dept. of Astronomy, Columbia University, 550 W 120th Street, New York NY 10027}

\date{Accepted . Received ; in original form }

\pubyear{2016}

\begin{document}
\label{firstpage}
\pagerange{\pageref{firstpage}--\pageref{lastpage}}
\maketitle

\begin{abstract}
	
The upcoming \TESS\ mission is expected to find thousands of transiting planets
around bright stars, yet for three-quarters of the fields observed the temporal
coverage will limit discoveries to planets with orbital periods below 13.7 
days. From the \Kepler\ catalog, the mean probability of these short-period
transiting planets having additional longer period transiters (which would be 
missed by \TESS) is 18\%, a value ten times higher than the average star. In 
this work, we show how this probability is not uniform but functionally 
dependent upon the properties of the observed short-period transiters, ranging
from less than $1$\% up to over $50$\%. Using artificial neural networks (ANNs)
trained on the \Kepler\ catalog and making careful feature selection to account 
for the differing sensitivity of \TESS, we are able to predict the most likely
short-period transiters to be accompanied by additional transiters. Through 
cross-validation, we predict that a targeted, optimized \TESS\ transit
and/or radial velocity follow-up program using our trained ANN would
have a discovery yield improved by a factor of two. Our work enables a 
near-optimal follow-up strategy for surveys following \TESS\ targets for 
additional planets, improving the science yield derived from \TESS\ and 
particularly beneficial in the search for habitable-zone transiting worlds.

\end{abstract}


\section{Introduction}
\label{sec:intro}

In 2013, NASA's \textit{Kepler Mission} ended its four year vigil of a 100 
deg$^2$ patch of the sky. The data obtained during this time will 
likely continue to be a rich vein of scientific discovery for years to come.
Whilst the \Kepler\ data has certainly revealed remarkable individual 
discoveries (e.g. \citealt{doyle:2011}; \citealt{darkest:2011}; 
\citealt{rappaport:2012}; \citealt{muirhead:2012}), the statistical insights
afforded by this homogenous catalog of over four thousand transiting planet
candidates have arguably been the most transformative (e.g. see 
\citealt{howard:2012}; \citealt{dong:2013}; \citealt{petigura:2013}; 
\citealt{dfm:2014}; \citealt{dressing:2015}; \citealt{burke:2015};
\citealt{traub:2016}).

The next major space-based transit survey will be NASA's \textit{Transiting
Exoplanet Survey Satellite}, or \TESS, expected to be launched 
late-2017/early-2018 \citep{ricker:2015}. Unlike \Kepler, \TESS\ will survey
a large fraction of the sky, seeking planets around the nearest and brightest
stars suitable for detailed subsequent characterization. The \TESS\ survey 
strategy comes at the cost of having to periodically shift fields to tile the
sky. This means that during the 2\,year nominal mission, more than 
three-quarters of the \TESS\ observed fields will be monitored for just 
27.4\,days. One major effect of this is that the ability of \TESS\ to discover 
long-period planets is severely diminished compared to that of the 
\KeplerMission\ \citep{sullivan:2015}.

Despite this, \TESS\ is expected to discover $\sim1700$ short-period transiting 
planets \citep{sullivan:2015}. Thanks to \Kepler, we know that multiple 
transiting planet systems are common, comprising $\sim20$\% of all observed
\Kepler\ systems \citep{coughlin:2016} and thus many of these ostensibly single
transiting planet \TESS\ systems will in fact have additional, longer-period
transiting planets missed by \TESS. Since only $\sim2$\% of \Kepler\ targets are 
observed to host transiting planets but $\sim20$\% of these have
multiple transiting planets, the $\sim1700$ short-period \TESS\ systems are much
more likely to be fruitful targets for subsequent transit searches than an
average target. Yet even with this advantage, conducting long-term precise
photometric monitoring of $\sim1700$ targets dispersed across the entire sky
would be non-trivial for ground- and space-based observatories.

In this work, we demonstrate that the probability of a short-period transiting
planet(s) (``inner(s)'') harboring an additional long-period transiting 
planet(s) (``outer(s)'') is not a single number, but instead is functionally 
dependent upon the properties of the short-period planet(s). This insight 
enables us to predict which single transiting planet systems are most likely to
be fruitful for subsequent transit follow-up. Applying ideas from machine 
learning, we train a feed-forward artificial neural network to the observed 
\Kepler\ catalog, which we show can be used to increase the yield of a mock 
transit follow-up program by a factor of two.

We briefly introduce artificial neural networks and our particular 
implementation in Section~\ref{sec:ANN}. Data preprocessing and training of the
network to the \Kepler\ sample is discussed in Section~\ref{sec:training}, 
including investigations of varying the properties of the network. We extend
our model to include an additional feature, yet still account for sensitivity
bias, using a hybrid network discussed in Section~\ref{sec:hybrid}. We discuss
the potential applications and physical interpretation of our work in 
Section~\ref{sec:discussion}.

\section{Artificial Neural Networks}
\label{sec:ANN}


Artificial neural networks (ANNs) are a class of machine learning techniques
designed to estimate or approximate complex functions, taking inspiration
from biological neural networks, such as those found in the brain. An 
ANN can be considered to be a function composed of many simple processing 
elements which relate an array of inputs, $\mathbfss{X}$, to an array of 
outputs, $\hat{\mathbfss{Y}}$. In what follows, we consider the case of 
\textit{supervised} ANNs only.

The $t^{\mathrm{th}}$ row of $\mathbfss{X}$ is a real-valued vector, 
$\bmath{x}_t$, of length equal to the number of different input variables,
$N$, where $t$ is the index of each training example. The different input 
variables are often called \textit{features}, which together describe the 
\textit{input pattern} to the network. Similarly, the $t^{\mathrm{th}}$ row of
$\hat{\mathbfss{Y}}$ is the corresponding, real-valued output vector, 
$\bmath{y}_t$, of length equal to the number of different output variables, 
$M$. The matrix $\hat{\mathbfss{Y}}$ is an approximation of the
desired output, $\mathbfss{Y}$.

Generally, an ANN is a structure of weighted interconnections between a layer
of input neurons, hidden layers of processing neurons and the final layer of 
output neurons (for example, see Figure~\ref{fig:network}). The hidden neurons
most often perform nonlinear scalar transformations (although they can also be
linear), described by their \textit{activation function}, $\Phi$. All inputs to
a neuron are multiplied by weights, often called \textit{synaptic strengths}, 
which are summed together and then transformed via the activation function 
\citep{haykin:1994}. The structure, number of hidden layers and neurons, as 
well as the activation functions, are chosen by the user and generally one aims
to use the simplest possible network to satisfactorily relate $\mathbfss{X}$ to 
$\hat{\mathbfss{Y}}$, in order to avoid over-fitting \citep{sarle:1995}. The 
synaptic strengths are then fitted to maximize the agreement between 
$\hat{\mathbfss{Y}}$ and $\mathbfss{Y}$.

\subsection{Feedforward Multilayer Perceptrons}
\label{sub:FF}

In this work, we have used one of the most commonly used types of supervised 
neural networks, that of \textit{multilayer perceptrons}, also commonly 
referred to as a \textit{feedforward} (FF) network (see \citealt{bishop:1995}
for an introduction). FF networks have been used in a variety of astronomical 
applications (for example, see \citealt{lundstedt:1994}; \citealt{snider:2001};
\citealt{vanzella:2004}; \citealt{graff:2014}) and are well-suited for 
classification problems \citep{bailer:2001}.

\begin{figure}
\begin{center}
\includegraphics[width=8.4 cm]{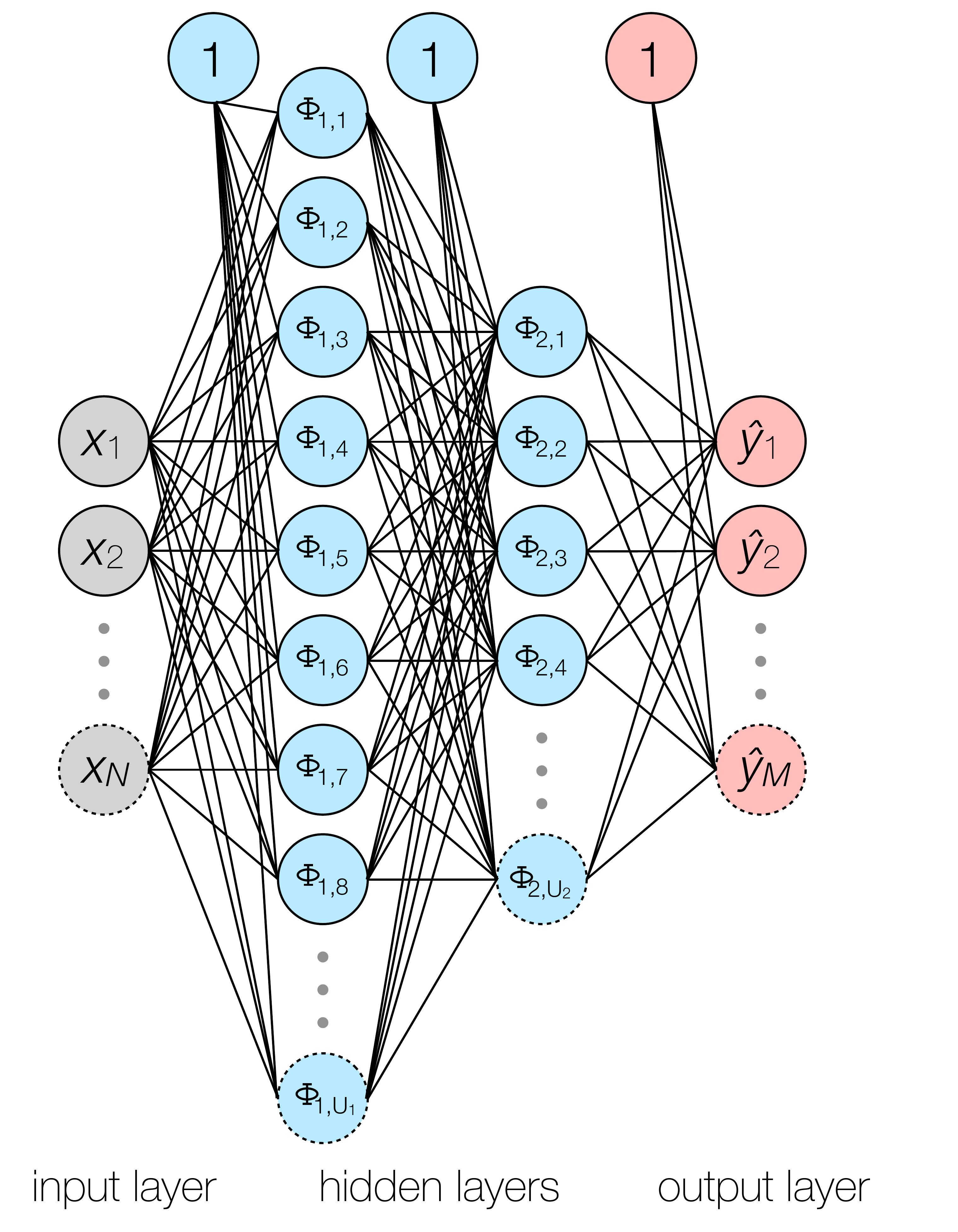}
\caption{
Architecture of a feedforward artificial neural network with two hidden layers.
An input pattern, $\bmath{x}$, is transformed into an output pattern, 
$\bmath{y}$, using the synaptic strengths, network structure, bias terms (1's)
and activation functions ($\Phi$).}
\label{fig:network}
\end{center}
\end{figure}

The FF network structure is illustrated in Figure~\ref{fig:network}, where we
depict two hidden layers although any arbitrary number is possible. For each
input pattern injected into the network, an output pattern is produced using 
the \textit{propagation rule}. Each hidden layer neuron is connected to all of
the previous layer's neurons, as well as a bias term, with variable synaptic
strengths. The output of the hidden neurons are then computed using the 
activation functions and fed-forward to the next hidden layer, which follows 
the same behavior. Whereas the hidden neurons typically use nonlinear 
activation functions, the output layer is calculated using the weighted sum of
the previous layer's output. In a classification problem, such as our own, this
output represents the class probability which may be converted to a binary 
classification using a logistic sigmoid.

Using this feedforward propagation, we may define the output pattern from the
FF network, $\bmath{y}$, given an input pattern, $\bmath{x}$. For example, the
$q^{\mathrm{th}}$ output neuron in a single-layer FF network, $\hat{y}_q$, will
be given by


\begin{align}
\hat{y}_q &= \beta_q + \sum_{k=1}^{U} v_{q,k} \Phi_k\Bigg[ \sum_{j=1}^N b_k + w_{k,j} x_j \Bigg].
\end{align}

The various terms in the above expression can be understood as follows.
$\beta_q$ is the synaptic strength to the output layer bias for the
$q^{\mathrm{th}}$ output neuron. $U$ is the number of neurons, or units, in
the hidden layer. $v_{q,k}$ is the synaptic strength between the
$q^{\mathrm{th}}$ output neuron and the $k^{\mathrm{th}}$ neuron in the
hidden layer. $\Phi_k$ is the activation function of the $k^{\mathrm{th}}$
neuron in the hidden layer. $b_k$ is the synaptic strength between the
hidden layer bias and the $k^{\mathrm{th}}$ neuron in the hidden layer.
$w_{k,j}$ is the synaptic strength between the $j^{\mathrm{th}}$ neuron
in the input layer of the $k^{\mathrm{th}}$ neuron in the hidden layer.

When training a single-layer FF network then, the vector of synaptic
strengths to fit, given by $\bmath{\theta} = \{ \bmath{\beta}, \bmath{u},
\bmath{b}, \bmath{w} \}$, has a total length of $\mathrm{dim}\bmath(\theta) 
= U + M + N U + U M$.

The above may be extended to more hidden layers, but requires introducing
extra notation indices for each layer, which we denote using subscripts.
The output of a double-layer FF network may now be expressed as


\begin{align}
\hat{y}_q &= \beta_q + \sum_{k=1}^{U_2} v_{q,k} \Phi_{2,k} \Bigg[ \sum_{j=1}^{U_1} b_{2,k} + w_{2,k,j} \Phi_{1,j} \Big[ \sum_{i=1}^N b_{1,j} + w_{1,j,i} x_i \Big] \Bigg].
\end{align}

Here, the total number of synaptic strengths is now $\mathrm{dim}\bmath(\theta)
= U_1 + U_2 + M + N U_1 + U_1 U_2 + U_2 M$.

\subsection{Activation Functions}
\label{sub:activation}

Whilst several non-linear activation functions are commonly used, the
rectifier function is the most popular in deep learning \citep{lecun:2015},
given by

\begin{align}
\Phi[z] &= \mathrm{max}(0,z).
\end{align}

Rectifiers are efficient to compute and differentiable (except at zero), yet do
not have the vanishing gradient problem afflicting traditional activation 
functions, such as the hyperbolic function \citep{hochrieter:1991}. For these
reasons, we elected to adopt this popular activation function in what follows
for all hidden neurons. We describe later how using an alternative activation
function does not affect the results of this work.

\subsection{Learning Algorithm}
\label{sub:learning}

In order to determine the synaptic strengths, we need to train the FF network.
First then, we must define a cost function to optimize.

In our work, the output layer comprises of a single neuron, which represents
the binary classification probability as to whether the input pattern, 
describing properties of a particular planetary system, has one or more 
additional transiting planets with periods $P>P_{\mathrm{cut}}$. The typical 
error function used is the mean square error (MSE), which for our problem is
given by

\begin{align}
\epsilon(\bmath{\theta}) &= \frac{1}{T} \sum_{t=1}^{T} (y_t - \hat{y}_t(\bmath{\theta}))^2,
\end{align}

where $T$ is the total number of training examples.

The most commonly used learning techniques for ANNs are back-propagation
algorithms, which essentially use some form of gradient descent to optimize
the cost function. In what follows, we elect to use one of the most popular
learning methods, the damped least squares Levenberg-Marquardt algorithm (LMA)
(\citealt{levenberg:1944}; \citealt{marquardt:1963}). We set the LMA to stop
once the cost function improves by less than 1 part in $10^8$.

We also update the cost function to include a modest regularization term, using
$L_2$\textit{-regularization}. Regularization is frequently used to help ANNs 
avoid overfitting the training set, pulling the learning back from fitting 
small noisy spikes and imposing a preference for an overall smoothness to the 
ANN function. This leads to our final cost function, $C$, of,

\begin{align}
C(\bmath{\theta}) &= \epsilon(\bmath{\theta}) + \delta \bmath{\theta}^T \bmath{\theta}
\end{align}

where $\delta$ is the $L_2$ regularization coefficient. After experimenting 
with different values, we settled on $\delta=0.1$ as providing a good balance 
between flexibility and smoothness. In order to validate the learning, we 
employ cross-validation and choose an FF network structure using the early 
stopping principle, both of which we describe later in Section~\ref{sub:cv} \& 
\ref{sub:earlystopping}, respectively.

\section{Training Data}
\label{sec:training}

\subsection{Training Data}
\label{sub:data}

We downloaded the \Kepler\ planetary candidates catalog from the
\href{http://exoplanetarchive.ipac.caltech.edu/}{NASA Exoplanet Archive}
\citep{akeson:2013} on May 17th 2016. Data were filtered such that only objects
dispositioned as planetary candidates by \Kepler\ were used, giving us 4696 
such objects.

We filtered the catalog for radii $R_P<32$\,$R_{\oplus}$, since objects larger
than this are very unlikely to be planets \citep{chen:2016}, and $\log g>4$,
since the false positive rate increases sharply beyond this 
\citep{sliski:2014}. These filters reduced our planet sample to 4022 planetary
candidates in 3056 systems.

\TESS\ can only strongly constrain the orbital period of planets which are
observed to undergo multiple transits. This means that only transiting planets 
with orbital periods less than $P_{\mathrm{cut}}=B/2$, where $B$ is the 
baseline of observations, are guaranteed to undergo $\geq2$ transits necessary
for a strong orbital period determination. For $\gtrsim75$\% of the \TESS\
observed fields, $B=27.4$\,d, since overlapping fields near the ecliptic pole
allow for greater $B$ \citep{sullivan:2015}. In this work, we adopt $B=27.4$\,d
and thus $P_{\mathrm{cut}}=13.7$\,d, although the ANN presented here could be
re-trained for other choices of $P_{\mathrm{cut}}$. After removing any systems 
for which there are no transiting planets with $P<P_{\mathrm{cut}}$, we are 
left with 1786 systems, each of which represents a training example for our 
ANN.

In each training example, we define a binary flag, $y_t=0$ or $y_t=1$, which
describes whether there are additional transiting planets with 
$P>P_{\mathrm{cut}}$ (or ``outers'') . In principle, \TESS\ will not have 
direct access to this information (except for the rare cases of single 
transits, which we ignore in this work). Whilst $y_t$ is not directly measured
with \TESS, our ANN aims to predict its likely value, based upon other features
which \TESS\ can observe.

\subsection{Features}
\label{sub:features}

We have some flexibility in how many inputs, or features, we wish to use for
the learning. Since we view our work mostly as an initial demonstration of the
power of ANNs to this problem, we seek to use a simple set of features rather
than performing an exhaustive search.

Guided by this principle, we note that the question we are asking, whether
there are \textit{additional} transiting planets, is predicated on the fact
there is one or more transiting planets with $P<P_{\mathrm{cut}}$ already
known to exist in each training system. Thus, for every training example, we
always have information relating to one or more pre-discovered short-period
transiters. The most basic parameters inferred from a transit are the orbital
period, $P$, and the planetary radius, $R$, making features related to these
terms obvious candidates for our ANN.

\begin{figure*}
\begin{center}
\includegraphics[width=16.6 cm]{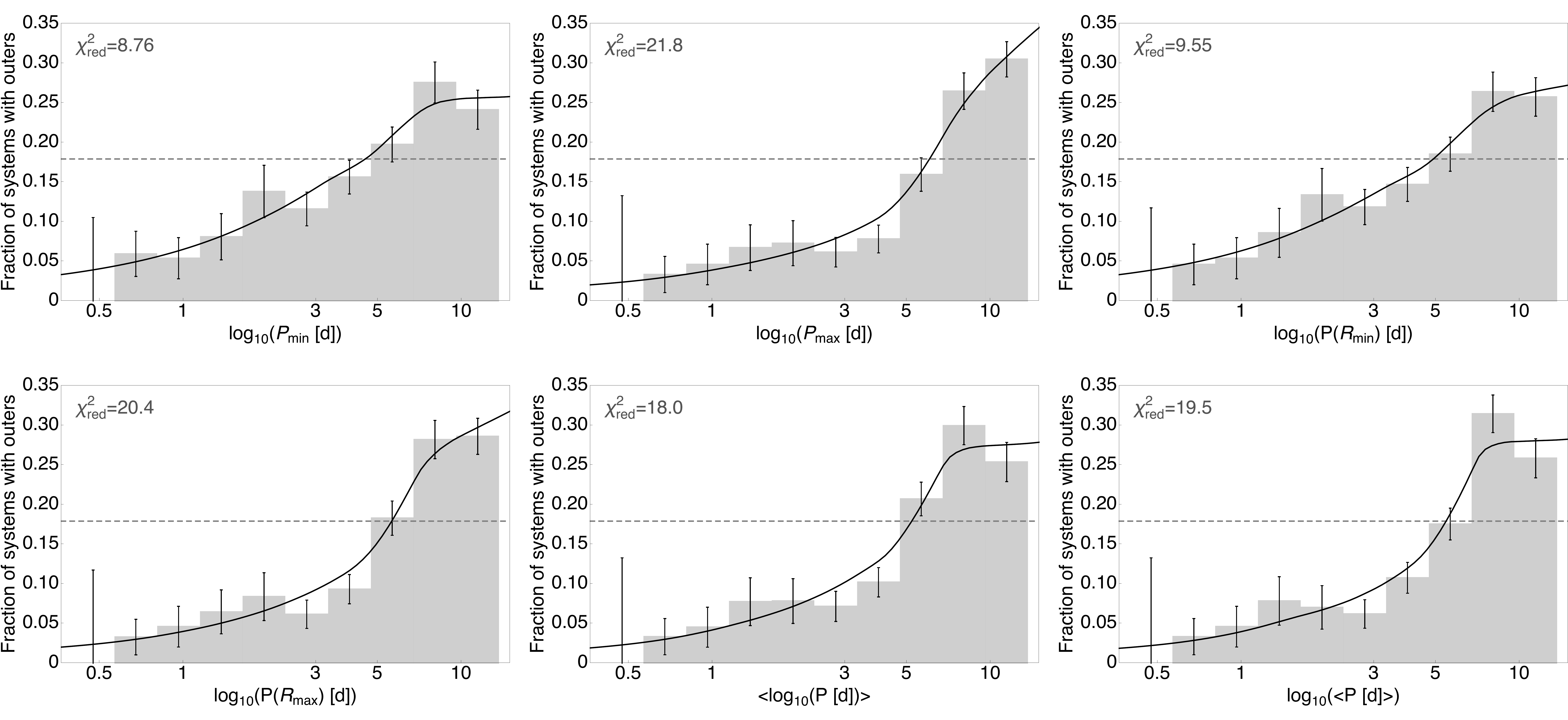}
\caption{
Initial exploration of the predictive power of each period-like feature.
Gray dashed horizontal line is the unconditioned a-priori probability of 
$0.179\pm0.009$. For comparison, the solid line shows the mean probability
predicted by 100 randomly initialized FF ANNs with one hidden layer of four
neurons, following the method described in Section~\ref{sec:ANN}.
}
\label{fig:Pfeats}
\end{center}
\end{figure*}

\begin{figure*}
\begin{center}
\includegraphics[width=16.6 cm]{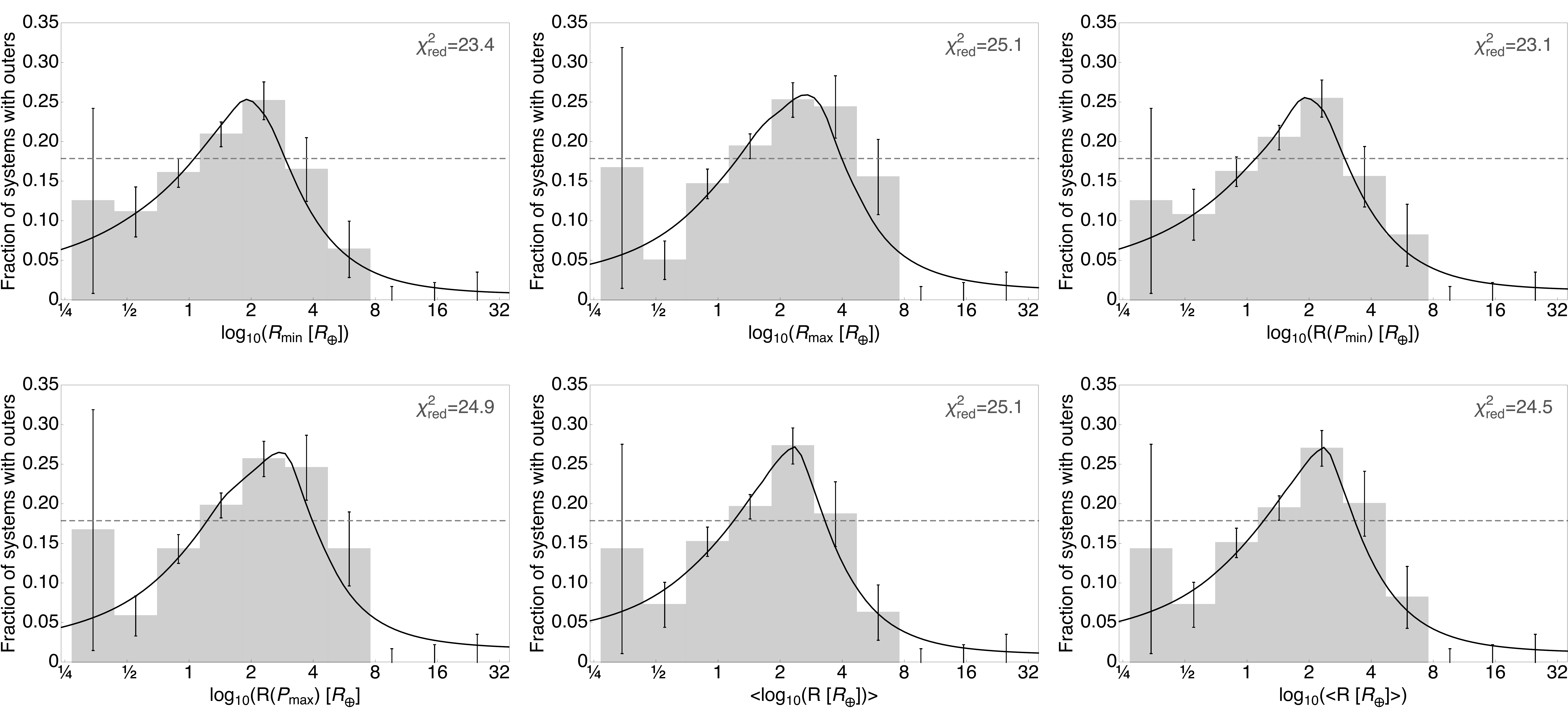}
\caption{
Same as Figure~\ref{fig:Pfeats}, except for radius-like features.
}
\label{fig:Rfeats}
\end{center}
\end{figure*}

We consider the $\log$ of each of these variables as potential features, since
they are observed to be more smoothly distributed versus linear space
\citep{dfm:2014}. For a system with $N_{\mathrm{inner}}$ transiting planets
with $P<P_{\mathrm{cut}}$ (or ``inners''), there are a number of single valued
features that could be constructed using period and radius, such as the mean,
minimum, maximum, etc logarithmic period/radius. In Figures~\ref{fig:Pfeats}
\& \ref{fig:Rfeats}, we show the fraction of outers observed as a function of
these various candidate features, where the gray line denotes the flat uniform
probability of the ensemble training set. The reduced $\chi^2$ deviance 
(computed using Poisson counting errors) of these fractions relative to the 
naive flat probability, which would be expected for irrelevant features, is
shown in the top-right of each panel, which can be treated as a feature
importance metric.

As evident from Figure~\ref{fig:Pfeats}, all of the period-like features show
similar trends to one another, and this is even more true in the case of the
radius-like features in Figure~\ref{fig:Rfeats}. We ultimately elect to use
$\log R_{\mathrm{max}}$ as our radius-like feature and 
$\log P(R_{\mathrm{max}})$ as our period-like feature since they are both
amongst the strongest features via the $\chi^2$ metrics, but more importantly
because they will generally always be available to us as features, for the
following reasons.

Whilst \Kepler\ and \TESS\ are both sensitive to small planets, \Kepler's much
larger aperture and staring time per field is expected to give it a greater 
sensitivity to the smallest sized planets \citep{sullivan:2015}. For this 
reason, the smallest sized planet in a systems with $N_{\mathrm{inner}}>1$ may
not be detectable by \TESS, even though it resides in our 
\Kepler-derived training set. In contrast, features related to the 
largest-sized planet will very likely always be available. This is because if
\TESS\ does not discover the largest-sized transiting planet with $P<13.7$\,d,
it will very likely not detect any transiting planets in that system at all, 
since transit detectability is primarily driven by the size of the planet 
\citep{sandford:2016}. Further more, if no transiting planets are known within
$P<P_{\mathrm{cut}}$, then the question we seek to answer, is a known 
transiter accompanied by additional transiting outers, is invalid since there 
are no known transiters to begin with.

\begin{figure*}
\begin{center}
\includegraphics[width=16.6 cm]{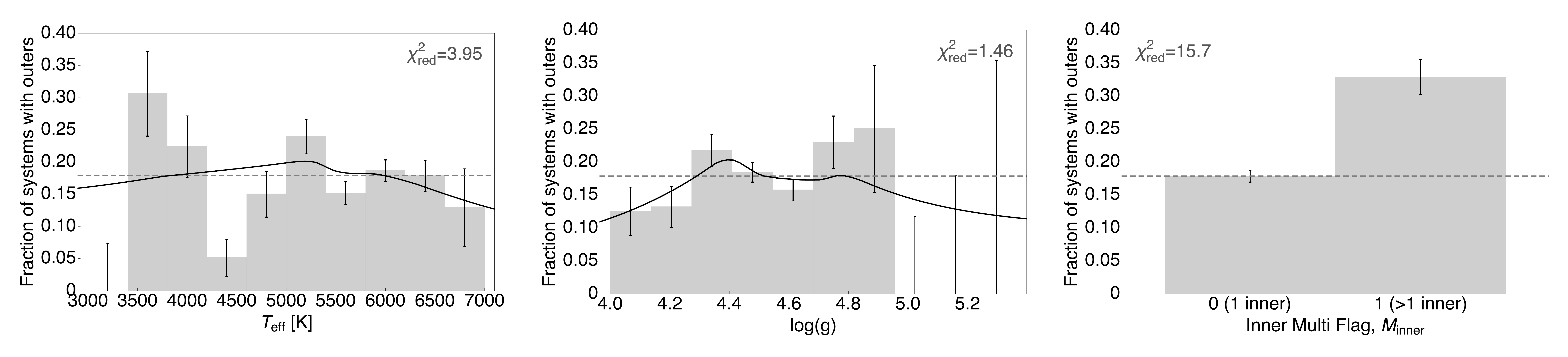}
\caption{
Same as Figure~\ref{fig:Pfeats}, except for three other features related to the
system.
}
\label{fig:Ofeats}
\end{center}
\end{figure*}

Having established $\log R_{\mathrm{max}}$ and $\log P(R_{\mathrm{max}})$ as
being viable and influential features, we briefly considered using other
features. Notably, we considered whether the properties of the star itself
could affect the probability of outers. Both the stellar effective
temperature, $T_{\mathrm{eff}}$, and surface gravity, $\log g$, appear to have
little predictive power with relatively low $\chi^2$ metrics, as shown in 
Figure~\ref{fig:Ofeats}. Consequently, we did not further consider stellar
parameters as potential features.

The final feature we considered was an inner multiplicity flag, which 
takes the value

\begin{align}
M_{\mathrm{inner}} &= \mathrm{min}(N_{\mathrm{inner}}-1,1).
\label{eqn:multi}
\end{align}

This feature clearly has a major influence (see Figure~\ref{fig:Rfeats}),
yet again it is a feature that may not always be available to us with
\TESS. Whilst the training set, based on \Kepler, is sensitive to small
planets, \TESS\ may miss them and again this feature is not generally
robust due to the different sensitivities between the two missions.
We therefore initially elect to ignore this feature and adopt a two-feature
input pattern of $\log R_{\mathrm{max}}$ and $\log P(R_{\mathrm{max}})$,
although we develop a hybrid model to include this feature later in 
Section~\ref{sec:hybrid}.

\subsection{Cross-Validation}
\label{sub:cv}

In most applications of ANNs, \textit{cross-validation} plays a critical role
in providing confidence in the accuracy and robustness of the ANN's
predictions, as well testing different ANN structures and parameters. The idea
behind cross-validation is to simply ignore a fraction of the available input
pattern, thereby treating the remaining data as the training set and the
excluded data as the validation set. The accuracy, however that is defined,
of the ANN when applied to the validation set provides a blind test of the
usefulness of the trained net. As described below, our treatment frames 
cross-validation in terms of how we envisage the ANN would be actually be used,
in order to blindly demonstrate its practical benefit.

Recall that our ANN will be used to predict which ostensibly single transiting
planets are most likely to harbor additional transiting planets at longer
orbital periods. Thus, the real world application of our ANN would be to input
the catalog of $\sim1700$ transiting planet candidates found by \TESS\ and
rank-order the most likely targets for subsequent, long-term photometric
monitoring. The purpose of our ANN is to optimize follow-up, such that it is
unnecessary to follow-up all of the \TESS\ candidates but rather some
fraction, $f$, selected by the ANN to most likely maximize the yield of new
discoveries. This line of thought provides a clear path to testing the value
of our trained ANN with cross-validation.

Cross-validation is performed by only training on 75\% of the 1786 training
examples comprising the \Kepler\ input pattern (see Section~\ref{sub:data} for
details). The remaining 446 training examples are treated as the validation
set. After learning has finished on the training set, we pass the validation
set through the ANN, which provides the class probabilities of each. We then
define some follow-up fraction, $f$, and after rank-ordering the objects by
their class probabilities, pick only the top $f$-quantile. From this follow-up
set, we calculate how many successes (i.e. additional transiting planets) were
actually observed, $S_{\mathrm{ANN}}$. We then repeat this same process except
the follow-up set is randomly selected, and the number of success is saved as
$S_{\mathrm{R}}$. The ratio of these values, $\mathcal{R}$, provides an 
estimate of how much better our ANN performs versus random picks, where

\begin{align}
\mathcal{R}=S_{\mathrm{ANN}}/S_{\mathrm{R}}.
\end{align}

\begin{figure}
\begin{center}
\includegraphics[width=8.4 cm]{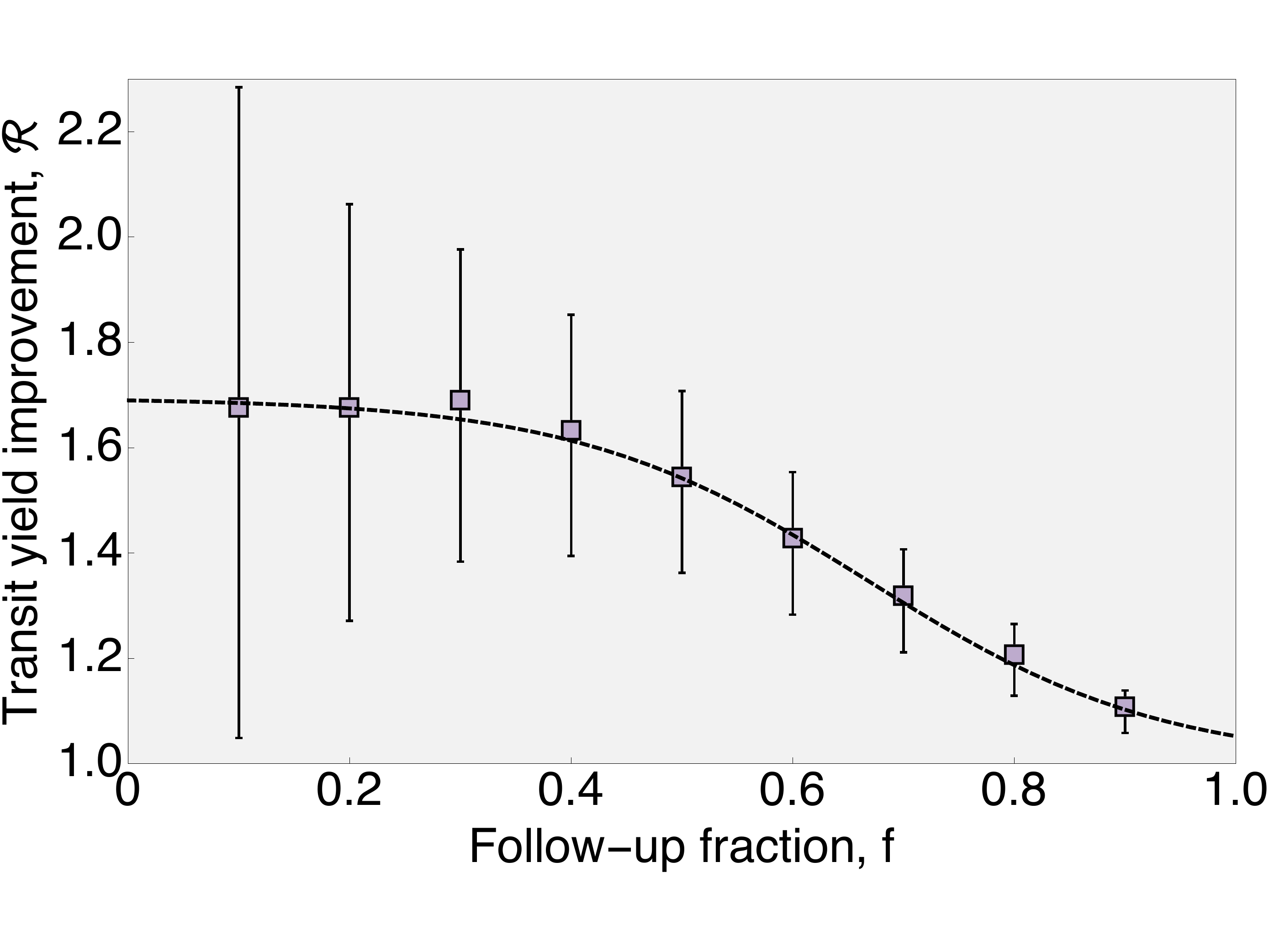}
\caption{
Cross-validation results for our ANN using a single hidden layer of $U=4$ 
neurons. We show the ratio of the number of successes achieved by our ANN 
versus random picks as a function of what fraction of the validation set is
followed-up, $f$. Points represent the median from $10^4$ Monte Carlo 
simulations, the errors represent the 68.3\% inter-quantile range and the
solid line is a sigmoid fit.
}
\label{fig:cv_example}
\end{center}
\end{figure}

Cross-validation is repeated $10^4$ times for any chosen $f$ value. The $10^4$
grid comprises of $10^2$ randomly sampled validation sets from the full set and
$10^2$ random initial seeds for the LMA. From the $10^4$ estimates of 
$\mathcal{R}$, we save the median and 68.3\% inter-quantile range
as our final estimates. We repeated the whole process for nine choices of $f$,
being 0.1, 0.2, 0.3, 0.4, 0.5, 0.6, 0.7, 0.8 and 0.9. As an example, we show
the resulting estimates of $\mathcal{R}$ as a function of $f$ in
Figure~\ref{fig:cv_example} for a single hidden layer FF ANN with $U=4$ 
neurons.

At $f=0.1$, just 45 validation samples are used, of which only
a fraction yield successes, leading to significant Poisson variance. At high
$f$, Poisson variance is suppressed but ultimately in the limit of $f=1$
any predictor cannot beat random pickings, since rank prioritization no longer
has any influence. We generated many example figures like
Figure~\ref{fig:cv_example} for different numbers of neurons, hidden layers
and activation functions and always see the same pattern of $\mathcal{R}$
saturating to some ceiling at $f\sim0.3$. We find that a logistic sigmoid
(solid line in Figure~\ref{fig:cv_example}) fitted through the cross-validation
results may be used to estimate

\begin{align}
\mathcal{R}_0 \equiv \lim_{f\to0} \mathcal{R},
\end{align}

which provides a single-valued metric to assess the cross-validation results
in what follows.

\subsection{Early Stopping}
\label{sub:earlystopping}

Properties of the FF ANN, in particular the number of hidden layers and number
of neurons, must be chosen. In general, one aims to use the simplest ANN which
can accurately relate the output pattern to the input pattern, in order to 
avoid overfitting and for computational expedience. One popular method for
choosing the ANN architecture is known as \textit{early stopping}, which we
use in this work. Essentially, the objective is to start from a simple model
and build up in complexity, stopping once the performance of cross-validation
no longer improves.

\begin{figure}
\begin{center}
\includegraphics[width=8.4 cm]{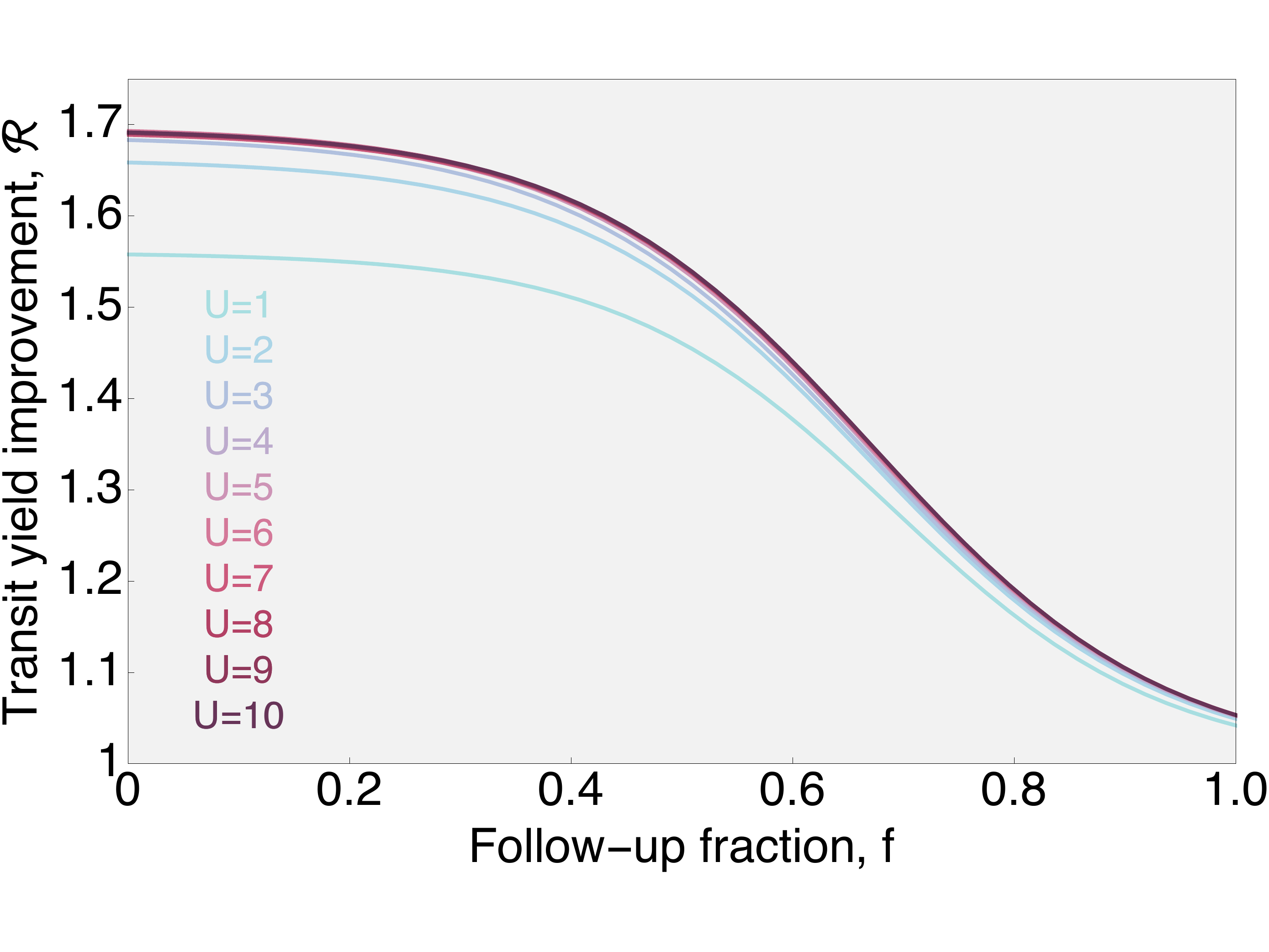}
\caption{
Cross-validation results from ten different single-layer FF ANNs, similar to
Figure~\ref{fig:cv_example} except we only show the fitted sigmoids. Using early
stopping, we identify that cross-validation performance does not improve for
$U>4$ and thus identify this as our preferred ANN.
}
\label{fig:early_stopping}
\end{center}
\end{figure}

Here, we start with a single-layer $U=1$ neuron FF ANN as our simplest model,
for which we find $\mathcal{R}_0=1.56$. The performance of even this very
simple ANN is impressive, yet notably lower than the $\sim1.7$ value achieved
with the single-layer $U=4$ ANN shown in Figure~\ref{fig:cv_example}.
Increasing $U$ up to 10, we find no improvement beyond $U=4$, which is evident
from Figure~\ref{fig:early_stopping}. Accordingly, we identify this model
as our preferred ANN, for which $\mathcal{R}_0=1.69$.

We repeated this exercise using a logistic sigmoid activation function
instead of the rectified linear function and find nearly identical results.
We also tried using a dual-layer network, exploring a variety of neuron
combinations up to a maximum complexity of $U_1=6$ and $U_2=6$,
but find $\mathcal{R}_0$ does not improve beyond $\simeq170$\%. As a final
test, we trained a triple-layer network with $U_1=10$, $U_2=8$ and $U_3=6$
and similarly found $\mathcal{R}_0=1.70$. These results imply that
a single-layer $U_1=4$ ANN is sufficient to capture the predictive power
of the selected features.

\section{Hybrid ANN}
\label{sec:hybrid}

\subsection{Multiple Inners Sample}
\label{sub:multis}

Of the full training set of 1786 systems, 307 (17\%) have multiple transiting
planets with $P<P_{\mathrm{cut}}$. Whereas the ensemble sample has a mean 
probability of hosting additional outers of $17.9\pm0.9$\%, this subset has a
much higher probability of $33\pm3$\% (as also shown in the right panel of 
Figure~\ref{fig:Ofeats}). In principle then, this multiplicity feature has a 
major influence on the class probability. But, as explained in 
Section~\ref{sub:features}, whilst the positive value of this feature could be
obviously detected by \TESS, the negative value cannot be ruled out, since 
\TESS's sensitivity is expected to be generally less sensitive to small planets
than \Kepler.

Nevertheless, if multiple inners are detected, the class probability of outer 
transiters is enhanced and can be calculated with an ANN. We therefore 
considered an additional ANN trained on three features, where the first two are
the same as before but the third is the multiplicity flag, $M_{\mathrm{inner}}$.

\subsection{Cross-Validation}

To cross-validate, we again emulate the practical way we envisage our trained
network being used. If a system is observed to have just one inner, we will
predict the class probabilities using the two-feature ANN from before, thereby
ignoring the inner multiplicity feature. The logic here is that these systems
may indeed have multiple inners, we just don't know it due to \TESS's
sensitivity bias, and thus we train on the ensemble set. If a system is
observed to have multiple inners by \TESS, then this would also be
true as observed in the training set derived from \Kepler. Accordingly, for
these instances we predict the class probability using an ANN trained using
the previously described three feature model.

Cross-validation is therefore identical to before except the output to the
network (and equivalently for the class probabilities) are now computed using:

\begin{align}
\hat{y}_t &=
(1-M_{\mathrm{inner}}) \hat{y}_t^{\mathrm{ANN2}} + 
M_{\mathrm{inner}} \hat{y}_t^{\mathrm{ANN3}},
\label{eqn:hybridclass}
\end{align}

where $M_{\mathrm{inner}}$ was defined earlier in Equation~\ref{eqn:multi},
and the superscripts ANN2 \& ANN3 denote the 2- and 3-feature ANN respectively.
We may now train ANN2 and ANN3 on a given training set, then use
Equation~\ref{eqn:hybridclass} to predict the class probabilities on the
associated validation set.

\begin{figure}
\begin{center}
\includegraphics[width=8.4 cm]{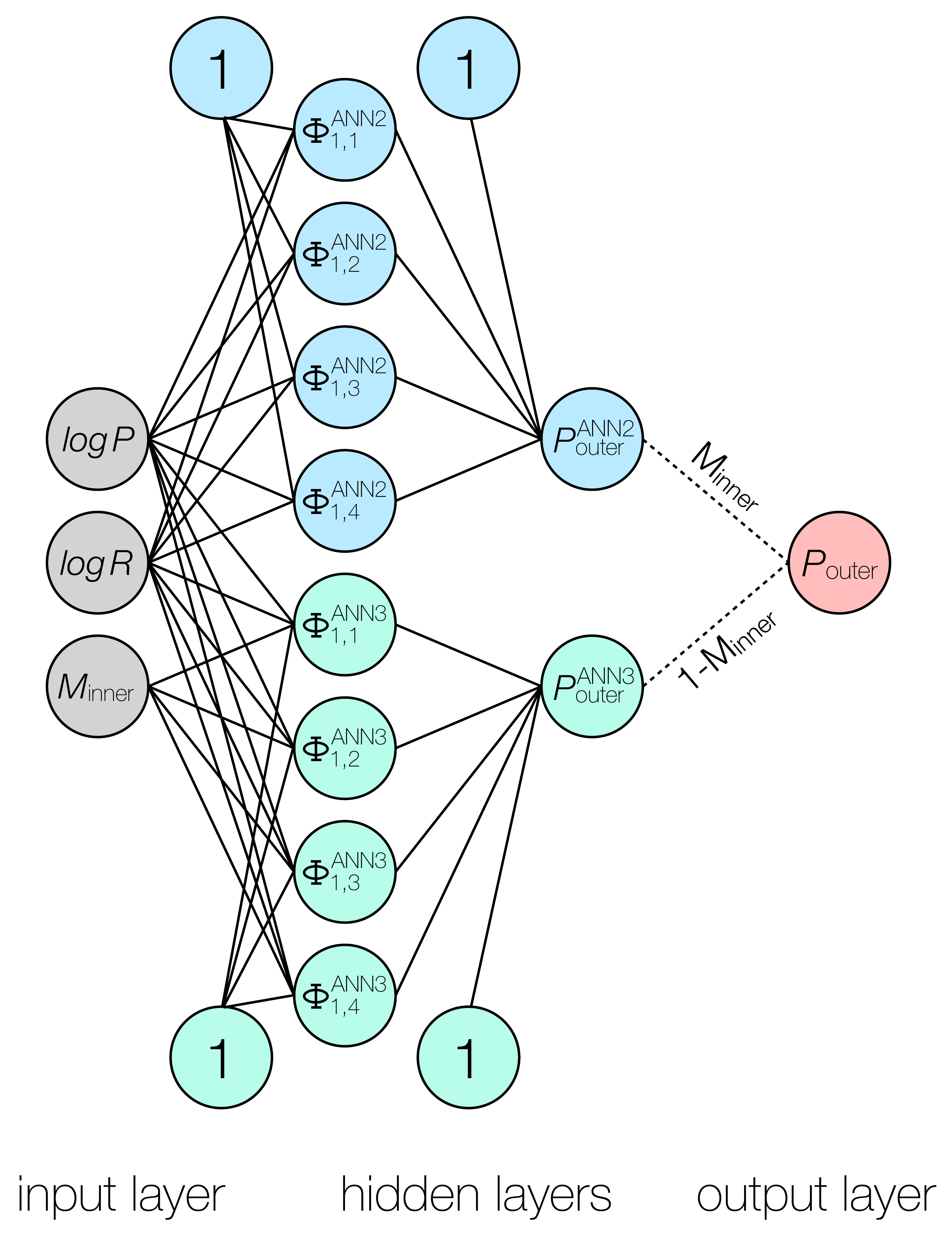}
\caption{
Architecture of our final network, which can be considered as being a hybrid
of two single-layer ANNs or one dual-layer ANN with several synaptic strengths
fixed to zero.
}
\label{fig:hybrid}
\end{center}
\end{figure}

In this way, we have constructed a hybrid of ANN2 and ANN3, which can be also
be thought of as a single ANN with a second hidden layer, comprising of
$\hat{y}_t^{\mathrm{ANN2}}$ and $\hat{y}_t^{\mathrm{ANN3}}$, and a first hidden
layer which has numerous synaptic strengths fixed to zero. The hybrid ANN 
structure is depicted in Figure~\ref{fig:hybrid}.

\subsection{Early Stopping}
\label{sub:hybridstopping}

As before, we use early stopping to identify the simplest possible ANN which
maximally improves the yield of transit surveys, described by parameter
$\mathcal{R}_0$. However, rather than varying the architectures of both ANN2
and ANN3 simultaneously, we fix ANN2 to the $U=4$ neuron preferred model found
earlier and explore single-layer, variable $U$ architectures for ANN3.

\begin{figure}
\begin{center}
\includegraphics[width=8.4 cm]{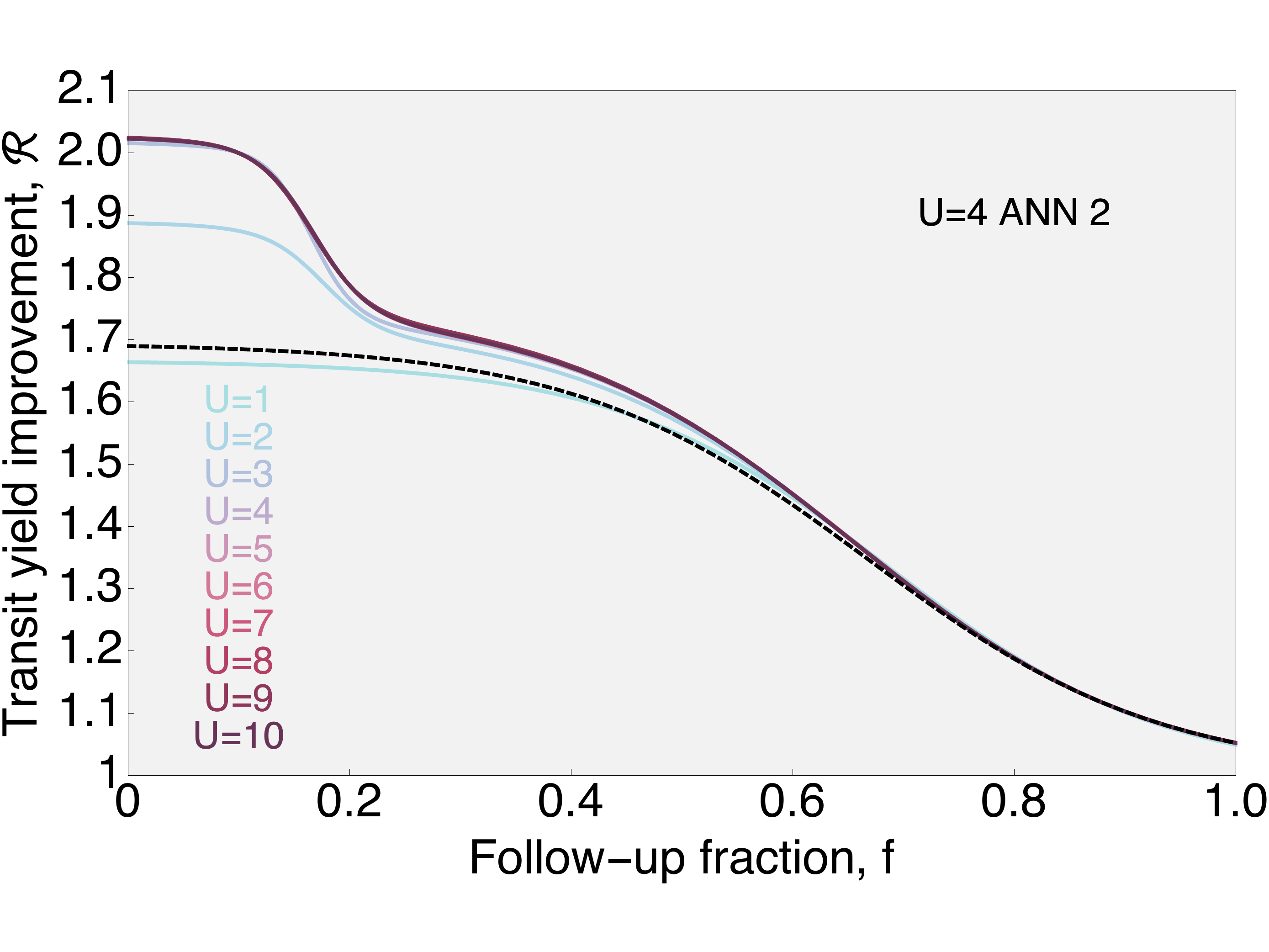}
\caption{
Cross-validation results from our hybrid 2/3-feature ANN, similar to
Figure~\ref{fig:cv_example} except we only show the fitted sigmoids. Using early
stopping, we identify that cross-validation performance stabilizes for
$U\geq4$ and thus identify this as our preferred ANN. For comparison, the black
dashed line shows the optimal $U=4$ ANN when $T_{\mathrm{eff}}$ is not used,
as found earlier in Section~\ref{sub:earlystopping}.
}
\label{fig:multis}
\end{center}
\end{figure}

As shown in Figure~\ref{fig:multis}, the cross-validation results do
not improve beyond $U=4$ and thus we select the four neuron architecture as the
preferred structure. Unlike the two-feature ANN, the cross-validation results 
display a steep change in $\mathcal{R}$ at $f\sim0.17$, except for the case of 
$U=1$ which again appears smooth. We model the results using two logistic 
sigmoids, extending upon the single logistic sigmoid used earlier in 
Section~\ref{sub:cv}.

As mentioned at the start of Section~\ref{sub:multis}, 17\% of the full
training set have $M_{\mathrm{inner}}=1$. Further, as evident from
Figure~\ref{fig:Ofeats}, these samples are nearly twice as likely to harbor
additional transiting outers. Therefore, the act of ranking the samples
from highest to lowest class probabilities and then selecting the best
$f=0.2$ quantile essentially defines a sample dominated by 
$M_{\mathrm{inner}}=1$ cases. The class probability is high for these cases,
but once we cross into $f\gtrsim0.2$, the top quantile starts to include 
$M_{\mathrm{inner}}=0$ samples, which have substantially lower class 
probabilities. This provides an explanation for the steep changes observed
in the cross-validation results of Figure~\ref{fig:multis}.

\section{Including Effective Temperature}
\label{sec:Teff}

\subsection{Overview}

The hyrbid ANN discussed in the previous section accounts for three features,
a radius-like feature ($R_{\mathrm{max}}$), a radius-like feature 
($P(R_{\mathrm{max}})$) and the inner multiplicity flag ($M_{\mathrm{inner}}$).
The feature selection was motivated by the feature exploration conducted in 
Section~\ref{sub:features} where properties relating to the star, namely
$T_{\mathrm{eff}}$ and $\log g$, appeared to have little influence on 
the class probability of an outer. Despite this, there are reasons to revisit
this choice.

In what follows, we focus on the feature $T_{\mathrm{eff}}$ since it displays
a higher $\chi^2$ than $\log g$ in the feature exploration conducted earlier.
Since giants have been removed from our sample, $T_{\mathrm{eff}}$ and $\log g$
both track the spectral type of the parent star and thus physically speaking
are proxies of the same thing.

The lack of a strong $\chi^2$ in Figure~\ref{fig:Ofeats} for $T_{\mathrm{eff}}$
does not necessarily imply it has no predictive power. Correlations with other
features could conspire together to mask the effect in a 1D marginalized format,
such as that presented in Figure~\ref{fig:Ofeats}. Moreover, it has been
established in previous works that the architectures of planetary systems
do vary as a function of spectral type (for example see 
\citealt{mulders:2015,dressing:2015}). However, we point out that these 
differences do not necessarily require that the specific output we train upon,
class probability of additional transiting planet beyond $P>13.7$\,days, will be 
substantially distinct.

\subsection{Simple Three-Feature ANN}

We decided to investigate whether including $T_{\mathrm{eff}}$ as an additional
feature improves the results of our ANN. We begin by taking the simpler 
(non-hybrid) ANN described in Section~\ref{sec:training}, which has just two
features ($R_{\mathrm{max}}$ \& $P(R_{\mathrm{max}})$) and augmenting it with
a third feature, $T_{\mathrm{eff}}$. Since the optimal architecture for the ANN
cannot be assumed to be the same as that of the previous two-feature case, we
repeated the exercise described in Section~\ref{sub:earlystopping} of comparing
the cross-validation detection yield improvement factor, $\mathcal{R}_0$, for
different ANN architectures.

\begin{figure}
\begin{center}
\includegraphics[width=8.4 cm]{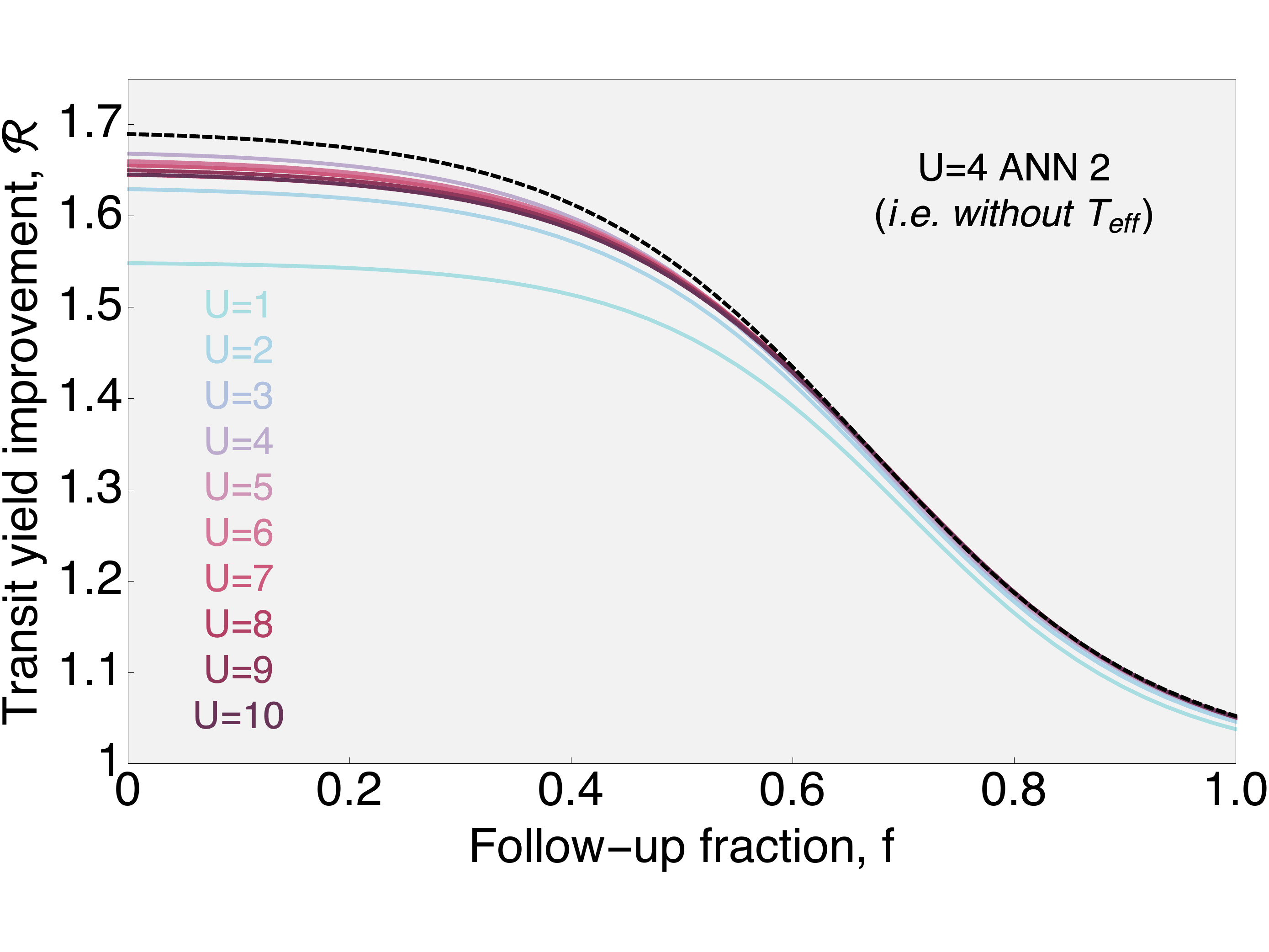}
\caption{
Same as Figure~\ref{fig:early_stopping}, except the single-layer FF ANN now
includes an additional feature, $T_{\mathrm{eff}}$. Cross-validation
actually degrades for networks using more than 4 neurons, indicating the
training is fitting out noise beyond this point. For comparison, the black
dashed line shows the optimal $U=4$ ANN when $T_{\mathrm{eff}}$ is not used,
as found earlier in Section~\ref{sub:earlystopping}.
}
\label{fig:Teff_early_stopping}
\end{center}
\end{figure}

Using a single-layer network, we varied $U$ from 1 to 10 and compared the 
$\mathcal{R}$ factors, as before. Unlike the two-feature case, the 
cross-validation results improve up to $U=4$ (as shown in 
Figure~\ref{fig:Teff_early_stopping}) but then start to worsen beyond
that beyond, which is implies we are starting to fit out noise which of course
has no predictive power.

Figure~\ref{fig:Teff_early_stopping} also reveals that including the third
feature does not include the cross-validation results, in fact it actually
degrades them. For example, the $U=4$ two-feature network gives
$\mathcal{R}_0=1.690$ whereas the $U=4$ three-feature network gives 
$\mathcal{R}_0=1.668$. We considered that perhaps this could be because the
network is not complex enough and thus tried using a $U=36$ single-layer 
network which gives $\mathcal{R}_0=1.678$, indicating that although the
predictions seem to improve somewhat from the trend seen in $U=5\to10$,
a single-layer ANN using $T_{\mathrm{eff}}$ is not able to recover a
prediction as accurate as the simple two-feature ANN can.

As before, we also investigated whether a dual-layer ANN could improve
the results and investigated all 36 combinations of $U_1$ \& $U_2$
from 1 to 6 (i.e. 36 total architectures). The best results were found
for the $U_1=4$ and $U_2=5$ network with $\mathcal{R}_0=1.669$, which
again is inferior even the single-layer two-feature model for $U=4$.
These results strongly support the result of the earlier exploratory
exercise that $T_{\mathrm{eff}}$ is not an influential feature in
predicting the existence of outers.

\subsection{Hybrid Four-Feature ANN}

To confirm this hypothesis, we also tried using the hybrid network described in
Section~\ref{sec:hybrid} but adding in $T_{\mathrm{eff}}$ as an extra feature 
once again. Since the previous subsection has established that the 
three-feature component of the hybrid model is optimized for $U=4$, we kept 
this part of the network fixed and explored different architectures for the 
component including $M_{\mathrm{inner}}$, as before.

As was found earlier, increasing $U$ beyond $U=4$ for this augmented network 
did not lead to any further improvements in the cross-validation results. Once
again, the inclusion of $T_{\mathrm{eff}}$ as an extra feature did not improve
the predictions of the hybrid network. The highest $\mathcal{R}_0$ value 
achieved was $\mathcal{R}_0=2.024$, nearly identical to the value found 
previously when $T_{\mathrm{eff}}$ was not included in Section~\ref{sec:hybrid}
($\mathcal{R}_0=2.024$).

From these results, we conclude that $T_{\mathrm{eff}}$ is not a useful feature
for predicting the existence of outer transiting planets in known transiting 
systems with $P<13.7$\,d. In all tests, networks without $T_{\mathrm{eff}}$ 
perform as well or better than those which include it. We stress that this 
result should not be interpreted as evidence that there are no significant 
differences between the architectures of planets orbiting stars of differing 
spectral types. Rather, it only speaks directly to the predictive power of this
very specific output that we have defined.

\section{Discussion}
\label{sec:discussion}

\subsection{Transit Clairvoyance}
\label{sub:summary}

In this work, we have demonstrated how artificial neural networks (ANNs) may be
used to predict which ostensibly single-planet short-period transiting systems 
are most likely to harbor additional longer-period transiting planets. 
Focussing on the upcoming \TESS\ mission \citep{ricker:2015}, for which in most 
fields the longest period transiters will be 13.7\,days \citep{sullivan:2015}, 
we show how follow-up photometric monitoring for additional transiters can 
expect to have the survey yield improved by a factor of two using our ANN (see 
Section~\ref{sub:hybridstopping}).

Although \TESS\ fields near the ecliptic pole overlap, permitting for
longer-period transit detections, the three-quarters of the survey fields
limited to 13.7\,day periods severely affects the ability of \TESS\ to detect
cooler, habitable-zone worlds. From our ANN, we find that some of the
short-period transiters have up to a $52$\% probability of hosting an
additional longer-period transiting planet (see Figure~\ref{fig:visualize})
though, providing an excellent opportunity to increase the science yield of 
\TESS\ and discover many more habitable-zone planets through ANN-guided 
targeted photometric follow-up from either the ground (e.g.
HAT \& HAT-S, \citealt{bakos:2004,bakos:2013};
KELT \& KELT-S, \citealt{pepper:2004,pepper:2012};
MEarth \& MEarth-S, \citealt{irwin:2009,irwin:2015};
NGTS, \citealt{wheatley:2013}; MINERVA, \citealt{swift:2014})
or space (e.g. MOST, \citealt{croll:2007}; CHEOPS, \citealt{broeg:2013}).

Detecting long-period transiting planets from the ground is challenging, 
due to the limited $\sim8$\,hour nightly observing windows. Ground-based
networks with longitudinal coverage, such as LCOGT \citep{brown:2013},
are designed to partially remedy this issue though. Naturally, the 
aforementioned space-based observatories do not suffer this limitation either.
Perhaps, though, the most fruitful follow-up program of these predicted planet
would not with photometry but via doppler spectroscopy. Here, the fact the 
outer planet is expected to be transiting maximizes the $m\sin i$ radial 
velocity amplitude yet observations would not have be to precisely timed to the
transit windows. Even a few sparse radial velocity points would be sufficient
to both confirm the presence of the planet and greatly narrow-down the likely
transit window, paving the way for a subsequent photometric detection.

Our approach should complement the alternative strategies of following up
\TESS\ single transit events \citep{yee:2008} and/or scheduling optimized
follow up of light curves which display no transits \citep{dzigan:2011}. In
the case of single events, long-period planets may fortuitously transit once
during the $B=13.7$\,day observing windows, with a probability given by $P/B$.
Our method complements this approach in that certain configurations of inner
transiting planet architectures have up to a 50\% probability of hosting
additional planets, yet may be quite unlikely to display a fortuitous
single transit during the observations.

To aid observers planning follow-up using our predictions, we make a grid of
the class probabilities with a \python\ example call available at
\wwwcoolworlds. 

Although our ANN only predicts the binary existence of outers,
we can compute an a-posteriori distribution of the period and radius of the
innermost outer (which is generally the easiest to blindly detect). To do this,
we take the full training data and extract the properties of the minimum period
outer, where they exist. The \TESS\ target stars do differ from
the \Kepler\ targets though, in particular the results of \citet{sullivan:2015}
show the effective temperature of planet hosting stars will be bimodal,
broadly defined as a mixture model with one component resembling the \Kepler\
sample and another peak at mid M-dwarfs. If the planet period and radius
distribution differs as a function of spectral type, our posterior would be
biased. To account for this, we simply exclude the stars in our sample
with $T_{\mathrm{eff}}<4450$\,K (although the effect of this is minimal on the
posterior). Accordingly, our posterior can be considered as being
conditioned on the fact not only an inner is detected but also 
$T_{\mathrm{eff}}>4450$\,K. Even so, we recommend this posterior only be
used as an approximate tool for predicting the parameter space of interest.

\begin{figure}
\begin{center}
\includegraphics[width=8.4 cm]{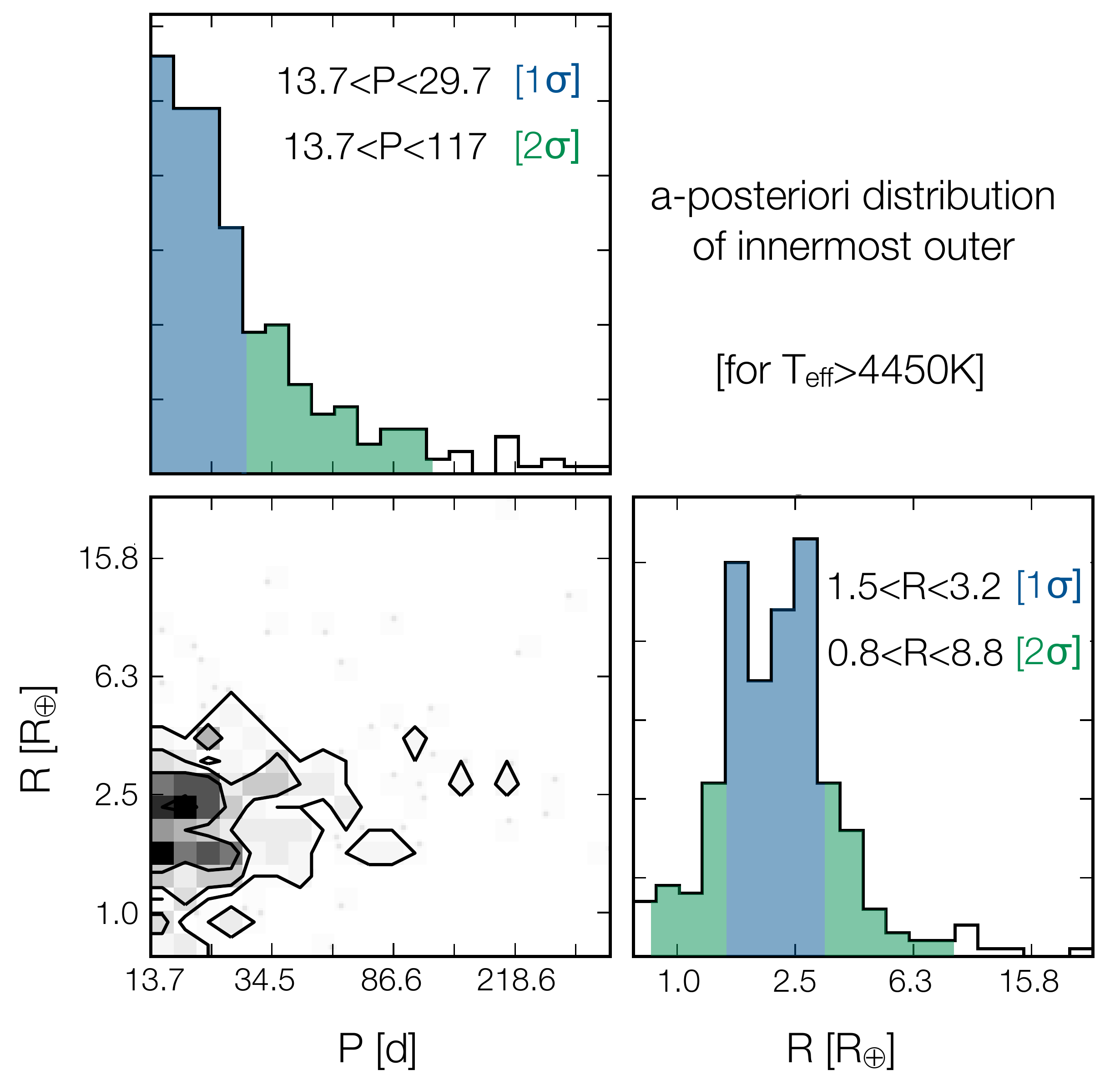}
\caption{
Triangle plot of the a-posteriori probability distribution of the innermost
outer transiting planet predicted from our ANN, conditioned on the fact
the host star has $T_{\mathrm{eff}}>4450$\,K.
}
\label{fig:posterior}
\end{center}
\end{figure}

The resulting joint posterior is shown in Figure~\ref{fig:posterior}, which can
be treated as an approximate prediction for the properties of outer planets 
using our ANN. To summarize, there is an 84\% chance of the period being below 
50\,d with the most likely size being 2.2\,$R_{\oplus}$.

\subsection{\Kepler\ vs \TESS}
\label{sub:KepvTESS}

Our ANN has treated the \Kepler\ catalog as the training set, although we
envisage the actual application to be on the \TESS\ catalog. One might
reasonably ask whether there are different sensitivities and biases at
play between the two, which may invalidate the results presented here.
Although the \TESS\ biases are not fully known yet, since the mission has
not flown, we argue it is unlikely that the differences between the
missions would invalidate the results here, by virtue of how we cognizantly
selected our features.

Both missions are essentially photon buckets optimized for approximately
visible bandpass photometry and seek planets in the same way. The major
differences, in terms of sensitivity, are that i) \Kepler\ stared at each
star for longer (4.35\,yrs versus 27.4\,d for most \TESS\ stars) ii) 
\Kepler\ has a larger aperture (0.95\,m versus 0.127\,m) iii) \TESS\ will
target brighter stars ($V=4$-$12$ versus $V=9$-$15$). Over the window of
$P<13.7$\,d, both missions have nearly continuous photometry and thus the
period effect, point i), only serves to increase \Kepler's photometric
sensitivity by (approximately) the ratio of their baselines square rooted.
Points ii) and iii) also both primarily affect the sensitivity to small
planets, in opposite directions.

Put together, \Kepler\ is expected to have a modestly
better sensitivity to small planets \citep{sullivan:2015}, but over the
range of $P<13.7$\,d their detection biases should be very similar, albeit
offset. In Section~\ref{sub:features}, we discuss how this insight
motivated us to use features related to the largest inner transiting planet
only, since \TESS\ may not see the smaller objects. If \TESS\ does not
detect even the largest planet, then one would not be attempted to use our
ANN anyway, since the question it poses is predicated on the assumption of
at least one known transiting planet.

A final concern we note is with \TESS's much greater focus on M-dwarfs than
\Kepler. As studied in detail in Section~\ref{sec:Teff}, we are unable to
find any indication that the class probability of an inner having an outer
is at all influenced by $T_{\mathrm{eff}}$. Since giants were already
excluded in our training set, then this indicates that the class probability
of interest is no different for M-dwarfs than for the FGK counterparts.
Based off these tests and the available information, we argue that our ANN
should be able to successfully increase the follow-up yield of \TESS\ as
described, unless the underlying planet population properties greatly differ
from those observed by \Kepler.

\subsection{Physical Insights}
\label{sub:physics}

Although the primary objective of this work is to predict which \TESS\ systems
are most likely to harbor additional transiting planets, we briefly discuss
the physical significance of our results here. We begin by visualizing the
probability space recovered by the ANN.

\begin{figure*}
\begin{center}
\includegraphics[width=17.0 cm]{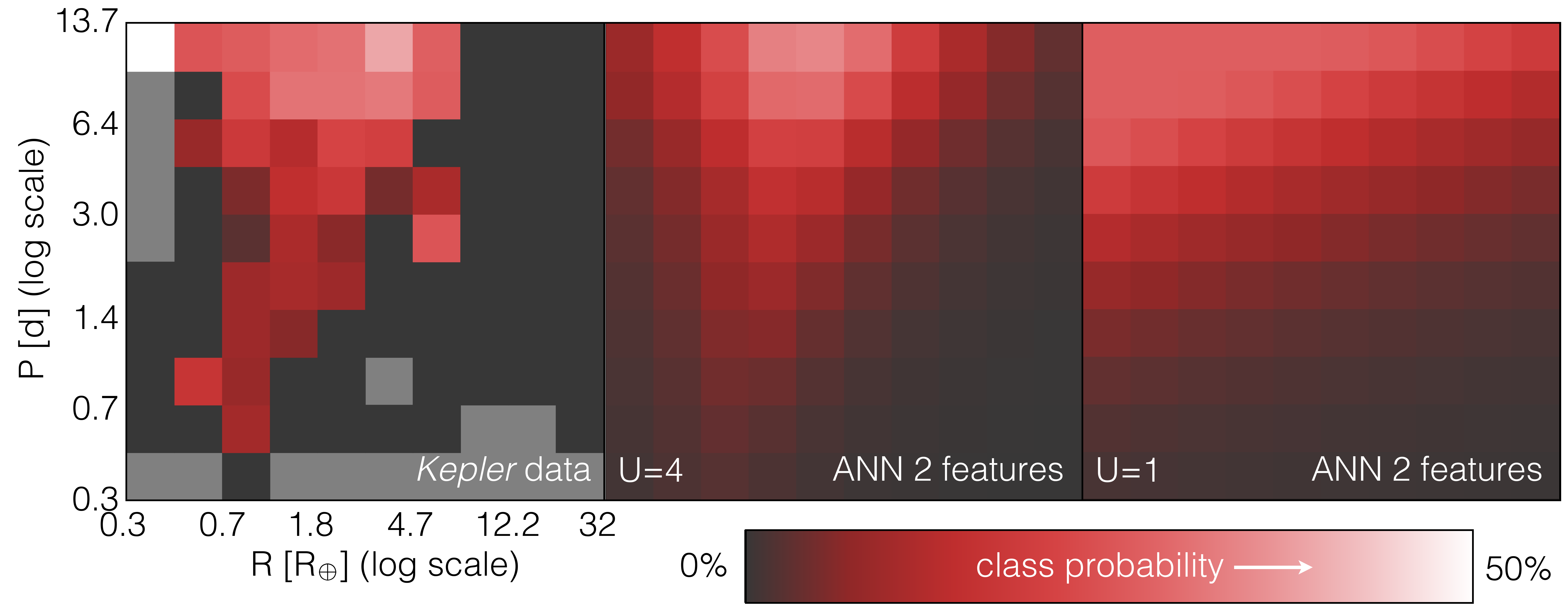}
\caption{
\textit{Left:} Fraction of short-period transiters with long-period transiting
companions, as observed in the \Kepler\ catalog and binned onto a 10 by 10
grid. Gray squares denote no data. \textit{Middle:} Class probabilities 
predicted by our preferred two-feature ANN (ANN2) with $U=4$ neurons, binned 
into the same space for a fair comparison. \textit{Right:} Same as middle, 
except using a simpler ANN with just a single neuron.
}
\label{fig:visualize}
\end{center}
\end{figure*}

In Figure~\ref{fig:visualize}, we show the probability of a known short-period
transiter having an outer as a function of $\log R_{\mathrm{max}}$ and
$\log P(R_{\mathrm{max}})$, as computed by our preferred two-feature ANN,
ANN2 (see middle panel). Specifically, ANN2 is trained on the entire training
set using our preferred structure (single-layer, four neurons; see 
Section~\ref{sub:earlystopping}) and the probabilities are averaged from $10^3$ 
random initial seedings of the LMA learning algorithm. We make a grid of these 
results available online at \wwwcoolworlds, which may be interpolated for 
arbitrary inputs. For comparison, we show the case of using a single-neuron ANN
in the right panel, which we find has inadequate flexibility. By binning the
input data onto a 10 by 10 grid, we can compare the results of the ANN to the
solution which it tried to learn (left panel).

Figure~\ref{fig:visualize} reveals a fairly monotonic probability space, with
a peak at long orbital periods and intermediate radii. We argue here that
the patterns observed at small $R$ are plausibly, although not unequivocally,
a result of the sensitivity drop-off of \Kepler\ in this region.

Firstly, \Kepler\ is known to have significant incompleteness for planets of 
$R\lesssim2$\,$R_{\oplus}$ \citep{christiansen:2016}. Second, even perfect
transit surveys have lower sensitivity to long-period transiters
\citep{sandford:2016} and in practice \Kepler's sensitivity to long-periods
drops even worse than this \citep{christiansen:2016}. Put together, this means
that the class probabilities at low $R$ can be lower than the truth, for
certain plausible planet populations. For example, if small inner planets tend
to be accompanied by approximately equal sized (or smaller) planets at longer
periods, then  whilst the inner one can be marginally detected by \Kepler, the
outer would evade detection despite its existence. This is compatible with
inferences made about the \Kepler\ population; in particular
\citet{ciardi:2013} find that for planets of $R\lesssim3$\,$R_{\oplus}$, there
is no correlation between the size and location of planets within a multiple
transiting planet system.

In contrast, the drop off from intermediate to large planetary radii is likely 
a real effect since completeness is very high here and only increases with
greater $R$. Jupiter-sized planets with periods below $P<13.7$\,d are usually 
considered to have migrated from beyond the snow-line \citep{lin:1996}, likely
disrupting the planetary system as they go \citep{steffen:2012,huang:2016}. 
These hot-Jupiters may therefore have either misaligned the other planets, such 
that they don't transit, or dynamically ejected them from the system. This 
picture is supported by the general trend seen in period space, where the 
shortest period planets rarely have outers.

\begin{figure}
\begin{center}
\includegraphics[width=8.4 cm]{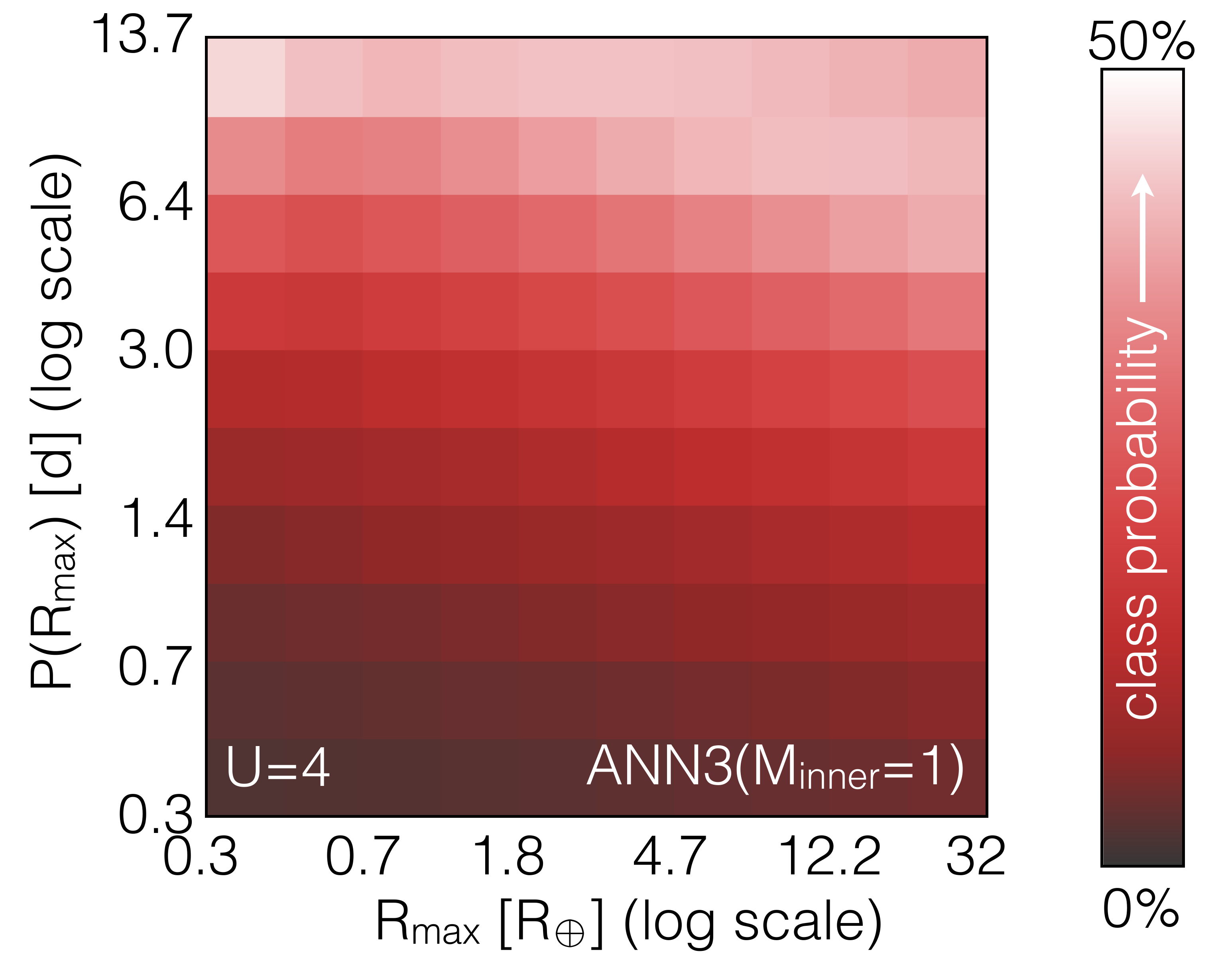}
\caption{
Same as middle panel of Figure~\ref{fig:visualize}, except we show the results
for the three-feature ANN (ANN3) when $M_{\mathrm{inner}}=1$. The class
probabilities are generally higher when more than one inner is detected and
become particularly high when the orbital period of the largest inner is high.
}
\label{fig:multisvis}
\end{center}
\end{figure}

In Figure~\ref{fig:multisvis}, we visualize the probability space again but
for the three-feature ANN, ANN3, in the case of $M_{\mathrm{inner}}=1$. Since
the output of ANN3 has zero-weight when $M_{\mathrm{inner}}=0$ (see Equation
~\ref{eqn:hybridclass}) in our hybrid network, those results have no bearing on
this work and thus are not shown. The ANN3 $M_{\mathrm{inner}}=1$ class 
probabilities are generally higher than that of ANN2, which is to be expected
based on the earlier feature investigations shown in Figure~\ref{fig:Ofeats}.
Additionally, the probability space is more uniform, likely as a result of much
sparser set of samples (recall that just 307 of the 1786 training samples have
$M_{\mathrm{inner}}=1$). However, there is a general preference for high
$\log P(R_{\mathrm{max}})$, suggesting that ultra compact systems are less
likely to have long-period transiting companions.

Together, these results are generally compatible with previously reported
trends in the \Kepler\ data, but we codify these trends with an ANN to enable 
quantitative predictions. This work provides one simple example application of
ANNs in exoplanetary science (see \citealt{waldmann:2016} for another recent
example) and we hope this introduction will motivate further applications of
this powerful machine learning technique to other problems in the future.
	
\section*{Acknowledgments}

We are grateful to the reviewer for their helpful suggestions and
constructive feedback.
We thank David Latham, Joshua Pepper, David Charbonneau and the Cool Worlds Lab
team for helpful comments and conversations in preparing this manuscript.
DMK acknowledges support from NASA grant NNX15AF09G (NASA ADAP Program).
This research has made use of the {\tt corner.py} code by Dan
Foreman-Mackey at 
\href{http://github.com/dfm/corner.py}{github.com/dfm/corner.py}.


%

\bsp
\label{lastpage}
\end{document}